\documentclass{article}

\usepackage{microtype}
\usepackage{graphicx}
\usepackage{subfigure}
\usepackage{booktabs} 
\usepackage{adjustbox}
\usepackage{comment}
\usepackage{verbatim}

\usepackage{amsmath,amssymb,amsfonts}
\usepackage{algorithm}
\usepackage[noend]{algorithmic}
\usepackage{textcomp}
\usepackage{xcolor}
 \usepackage{multirow}

\usepackage{hyperref}

\usepackage[accepted]{mlsys2020}

\mlsystitlerunning{Towards Latency-aware DNN Optimization with GPU Runtime Analysis and Tail Effect Elimination
}

\begin{document}

\twocolumn[
\mlsystitle{Towards Latency-aware DNN Optimization\\ with GPU Runtime Analysis and Tail Effect Elimination
}




\begin{mlsysauthorlist}
\mlsysauthor{Fuxun Yu}{to}
\mlsysauthor{Zirui Xu}{to}
\mlsysauthor{Tong Shen}{goo}
\mlsysauthor{Dimitrios Stamoulis}{goo}
\mlsysauthor{Longfei Shangguan}{goo}
\mlsysauthor{Di Wang}{goo}

\mlsysauthor{Rishi Madhok}{goo}
\mlsysauthor{Chunshui Zhao}{goo}
\mlsysauthor{Xin Li}{goo}
\mlsysauthor{Nikolaos Karianakis}{goo}
\mlsysauthor{Dimitrios Lymberopoulos}{goo}

\mlsysauthor{Ang Li }{ed}
\mlsysauthor{ChenChen Liu }{umd}
\mlsysauthor{Yiran Chen }{ed}
\mlsysauthor{Xiang Chen}{to}
\end{mlsysauthorlist}

\mlsysaffiliation{to}{George Mason University,}
\mlsysaffiliation{goo}{Microsoft,}
\mlsysaffiliation{ed}{Duke University,}
\mlsysaffiliation{umd}{University of Maryland, Baltimore County}

\mlsyscorrespondingauthor{Fuxun Yu}{fyu2@gmu.edu}

\mlsyskeywords{Machine Learning, MLSys}

\vskip 0.3in

\begin{abstract}
	Despite superb performance of State-Of-The-Art (SOTA)  DNNs, the increasing computational cost makes them very challenging to meet real-time latency and accuracy requirements. 
	Although DNN runtime latency is dictated by model property (\textit{e.g.}, architecture, operations), hardware property (\textit{e.g.}, utilization, throughput), and more importantly, the effective mapping between these two, many existing approaches focus only on optimizing model property such as FLOPS reduction and overlook the mismatch between DNN model and hardware properties. 
    In this work, we show that the mismatch between the varied DNN computation workloads and GPU capacity can cause the idle \textit{GPU tail effect}, leading to GPU under-utilization and low throughput.
	As a result, the FLOPs reduction cannot bring effective latency reduction, which causes the sub-optimal accuracy versus latency trade-offs.
Motivated by this, we propose a GPU runtime-aware DNN optimization methodology to eliminate such GPU tail effect adaptively on GPU platforms. 
    Our methodology can be applied on top of existing SOTA DNN optimization approaches to achieve better latency and accuracy trade-offs. 
    Experiments show 11\%-27\% latency reduction and 2.5\%-4.0\% accuracy improvement over several SOTA DNN pruning and NAS methods, respectively. 

\end{abstract}

]



\printAffiliationsAndNotice{}  

\section{{Introduction}}
\label{sec:intro}

Deep Neural Networks (DNNs) have achieved great performance on various cognitive applications, such as image classification~\cite{imgnet}, object detection~\cite{coco}, speech recognition~\cite{speech}, \textit{etc}.
	However, such a success is built upon a considerable cost of computing resources with increasingly larger model parameter volume and structure complexity.

To relieve the computation cost and improve system performance, many DNN optimization techniques have been studied.
	Started with an algorithm perspective, deep model compression once became a mainstream approach, such as the weight sparsity~\cite{hansong}.
	Although theoretically outstanding, these methods' effectiveness was latterly proved to be hardly translated into actual computation load reduction~\cite{harmful}.
	%
Further works also demonstrated that practical optimization has to consider the hardware perspective, and match the model structure reconfiguration with the hardware mechanisms (\textit{e.g.} structured pruning~\cite{lihao, prune1}).
	Therefore, the design gap emerged between the theoretical algorithm design and hardware deployment, which considerably complicated the current DNN development.

To fill this gap, many hardware-aware DNN designs have been proposed recently~\cite{net_adapt, proxylessnas}.
	One of the current masters is network architecture search (NAS).
	Leveraging various DNN configuration perspectives and methods, NAS profiles different DNN structures' performance on dedicated systems, in terms of floating point operations (FLOPs), overall latency, power consumption, \textit{etc.}~\cite{fbnet, mnasnet}.
	Such profiling is further utilized to search and identify the optimal DNN structure to accommodate specific system expectations.


However, hardware-aware algorithm designs as comprehensive as NAS, also have certain underlying ineffective performance interpretations, just like non-structured weight sparsity back then.
	On one hand, the understanding of the DNN structure's impact on the system performance is still based on ``end-to-end'' profiling, lacking the understanding of the intrinsic operations~\cite{proxylessnas, mnasnet}.
	In other words, the optimized DNN model configuration may not fully match the hardware mechanisms.
	On the other hand, FLOPs reduction is always utilized as a general performance indicator by previous optimization works~\cite{hrank, effnet}.
	However, the actual system gain translation between FLOPs and other performance perspectives is barely analyzed before, especially when applying such a static indicator into dynamic operations in terms of runtime latency.

\begin{figure}[!tb]
	\centering
	\includegraphics[width=3.3in]{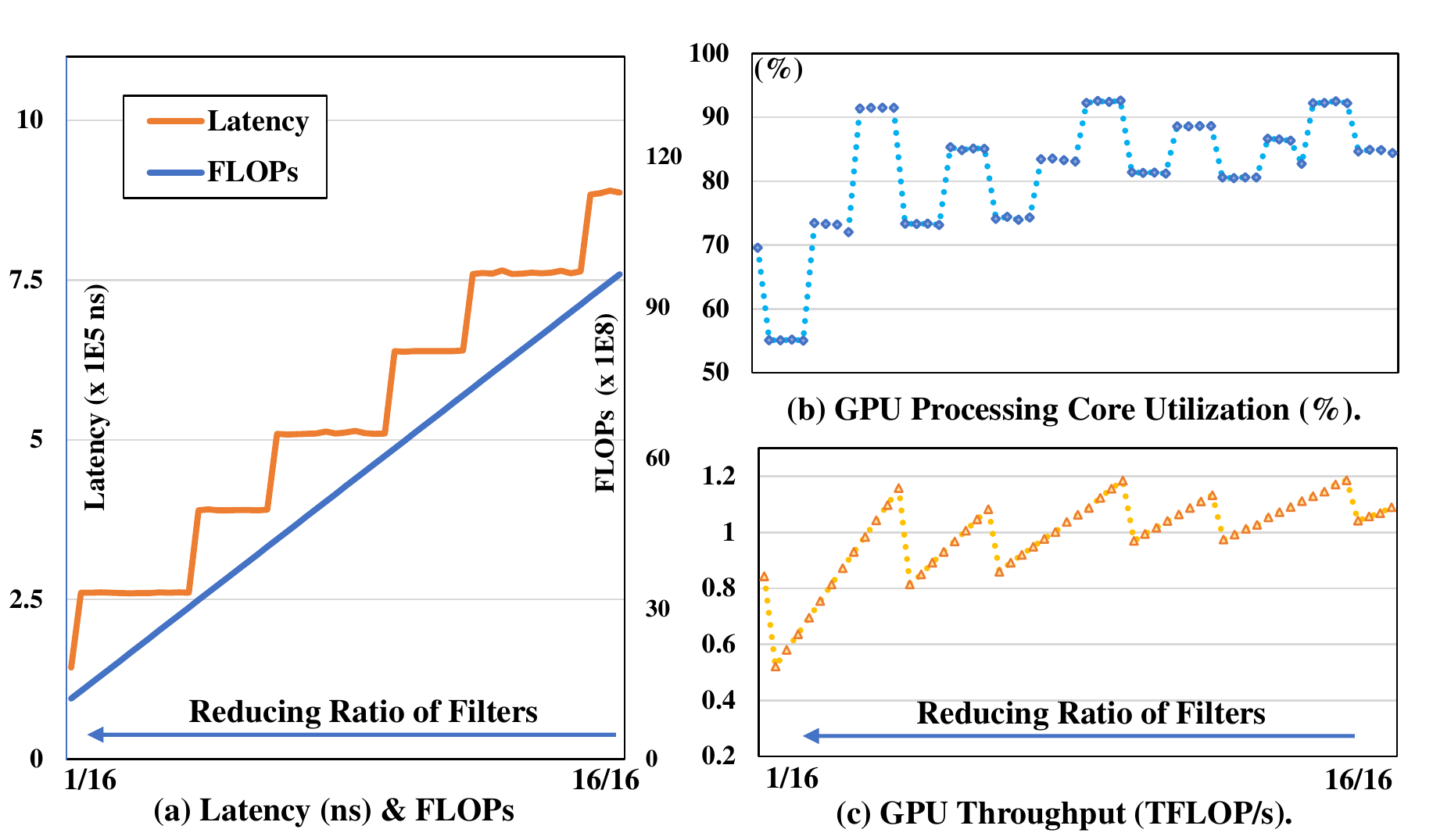}
	\vspace{-5mm}
	\caption{When reducing a convolutional layer's FLOPs by pruning filters, the layer's GPU runtime latency only reduces occasionally, forming a ``latency staircase'' pattern.
	}
	\label{fig:1}
	\vspace{-2mm}
\end{figure}

\textit{Our analysis verifies these issues from a runtime latency perspective, and shows that the FLOPs reduced by popular DNN optimization methods could not consistently translate to the runtime latency reduction in the practical system processing.}
Fig.~\ref{fig:1} shows the optimization results of a DNN model -- ResNet50 with a structural filter pruning method~\cite{lihao}.
	Specifically, one convolutional layer's FLOPs and latency reduction are plotted in Fig.~\ref{fig:1}(a).
	It is obvious that, although the FLOPs perform a linear reducing trend with constantly pruned filters, the latency reduction demonstrates a non-linear ``staircase'' shape.
	Such a \textit{latency staircase} indicates the inconsistency of the performance translation from FLOPs to latency reduction.
	Moreover, when we look into the GPU processors' utilization rate and throughput, as shown in Fig.~\ref{fig:1}(b)(c), they demonstrate a fluctuated decreasing trend but with certain cyclical patterns, which implies the potential GPU operation characteristics the adopted optimization cannot reach.

Motivated by such an observation, in this work, we focus on the practical GPU latency optimization with DNN processing.
	Although GPU runtime optimization has been well studied in conventional tasks, the aforementioned DNN related issues are still highly overlooked.
	Therefore, we explore the practical GPU runtime mechanism behind DNN processing, and propose a set of practical latency optimization solutions.
	Specifically, we make the following contributions:
	\vspace{-2mm}
\begin{itemize}
	\item Different from previous ``end-to-end'' DNN performance profiling, we conduct a ``full-stack'' analysis that dives into individual GPU processing cores and threads to examine DNN deployment and processing's intrinsic mechanism.
		By revealing the root cause of the latency staircase phenomena, we redefine the ``GPU tail effect'' for DNN computation from practical GPU utilization and throughput perspectives. (\S\ref{sec:3})

	\vspace{-1mm}
	\item We propose a set of GPU runtime-aware DNN model structure configuration methodology, which eliminates the GPU tail effect with thread-adaptive DNN deployment.
		The proposed methodology can also be utilized to enhance cutting-edge DNN optimization algorithms to escalate their system latency performance. (\S\ref{sec:4})

	\vspace{-2mm}
	\item Extensive experiments across common benchmarks are conducted.
		Specifically, the effectiveness of the proposed methodology is demonstrated 
		with distinguishable system speed up and accuracy enhancement, e.g., 11\%-27\% latency and 2.5\%-4.0\% accuracy improvement over SOTA DNN pruning and NAS methods.
		Moreover, the feasibility and generalizability of the methodology are also well discussed with various algorithms and hardware configurations. (\S\ref{sec:exp})
\end{itemize}

\section{{Background and Related Work}}
\label{sec:background}


\subsection{GPU Architecture Overview}
\label{ss:gpu_arch_overview}
As our work mainly focuses on DNN optimization on GPUs, we first describe the GPU architecture from both hardware architecture and computation deployment perspectives.

\begin{figure}[!tb]
	\centering
	\includegraphics[width=3.3in]{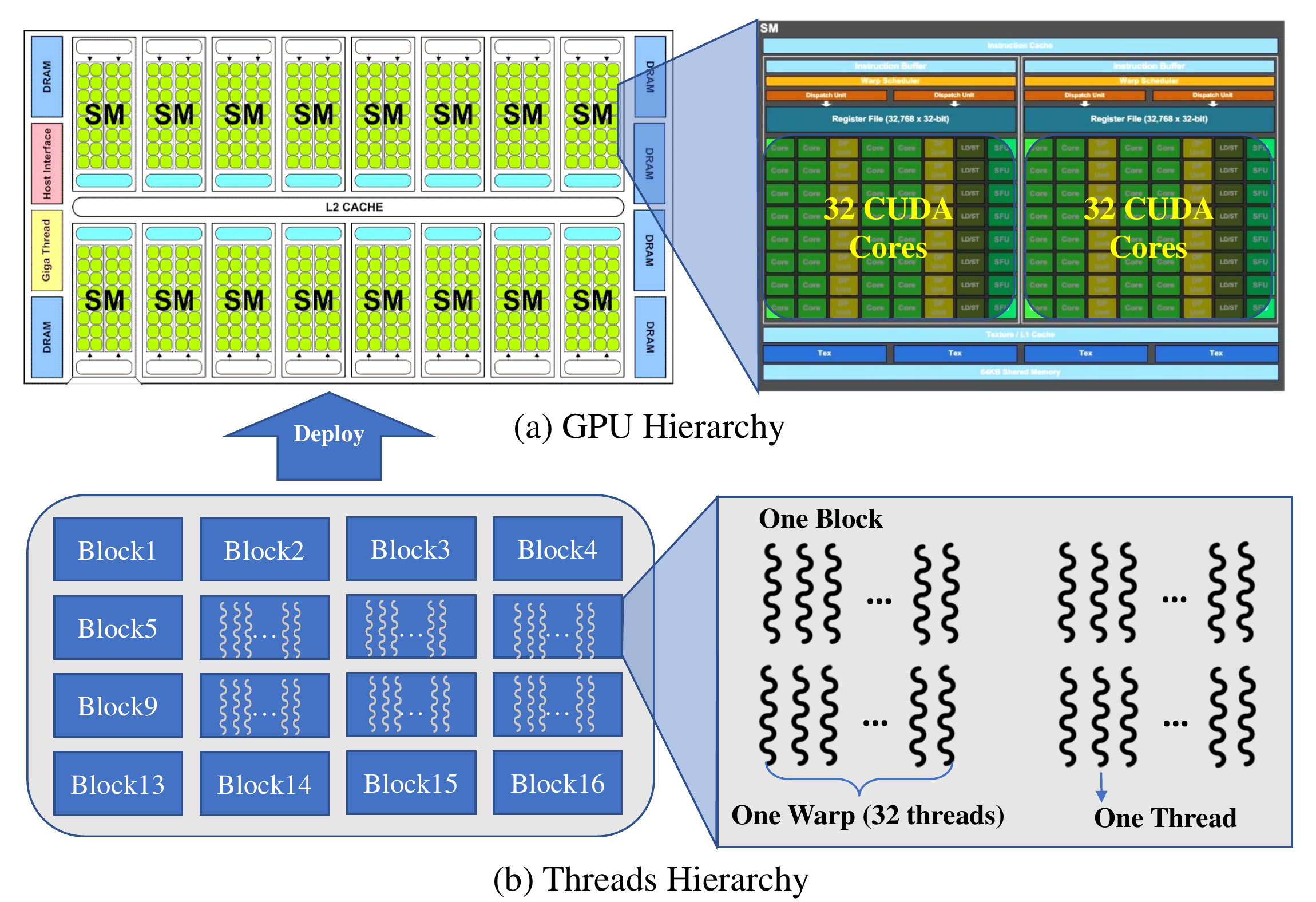}
	\vspace{-5mm}
	\caption{An illustration on parallel workload mapping to GPUs. (a) The GPU architecture hierarchy is composed of SMs and CUDA cores. (b) The parallel workload is also logically split into blocks and threads so as to map to the GPU hierarchy.}
	\vspace{-2mm}
	\label{fig:gpu}
\end{figure}

\paragraph{GPU Architecture Hierarchy}
Without loss of generalizability, we take NVIDIA architecture \cite{fermi} as an illustrative example. 
As shown in Fig.~\ref{fig:gpu}(a), the GPU is made up of  an array of {\it Streaming Multiprocessors (SMs)}, each of which is formed by multiple {\it CUDA Cores}.
For example, NVIDIA Titan-V GPU \cite{titanv} is made up of 5120 CUDA cores, which are grouped into 80 SMs.
Similar mechanisms also apply to other GPU architectures like AMD Polaris \cite{polaris}.

\paragraph{Computation Deployment into Threads}
When deploying a multi-threaded program into a GPU, the program will be partitioned into {\it Blocks} of threads for processing.
	Each thread block contains certain groups of {\it Warps}, and each warp composes of 32 {\it Threads}, as shown in Fig.~\ref{fig:gpu}(b).

\vspace{-0.5mm}
Therefore, thread blocks are the basic units deployed to SMs (one-to-one, or multi-to-one depending on the SM capacity).
    When the number of thread blocks exceeds the maximum GPU capacity, it would take multiple GPU processing cycles to complete the deployed workload.
	In such cases, multiple \textit{Waves} of thread blocks will be processed sequentially until the workload is completed.

\subsection{Runtime Latency on GPUs}
\label{ss:runtime_latency}
%
Runtime latency reflects the GPU execution efficiency given particular workload and throughput~\cite{vivie}:
\begin{equation}
	\vspace{-1mm}
	\text{Latency} = {\text{Workload}} / {\text{Throughput}}.
	\vspace{-1mm}
	\label{eq:1}
\end{equation}

When processing DNNs, the workload is quantified by the number of floating point operations, \textit{i.e.}, {\it FLOPs}.
{\it Throughput} measures the number of floating point operations per second (FLOPs/s).
	Taking a step further, when loading a DNN model into multiple SMs, the GPU throughput will be affected by $i)$ peak throughput per SM, $ii)$ number of SMs, and $iii)$ the \textit{utilization} of SMs:
\begin{equation}
	\vspace{-1mm}
	\centering
	\begin{split}
		\text{GPU Throughput} = \text{Peak throughput per SM} \\
		~\times~ \text{Number of SMs} \times \text{Utilization of SMs},
		\label{eq:2}
	\end{split}
	\vspace{-1mm}
\end{equation}

The peak throughput and the number of SMs are determined by the GPU hardware and thus can be regarded as constant values.
While the GPU {\it utilization} changes over time throughout the model execution~\cite{cuda}.

In a nutshell, both model-level workload and hardware-level GPU throughput (also SM utilization) affect the runtime latency of DNN models.
	However, most of the existing works focus only on the model-level workload optimization and omit the hardware characteristics, as we will show next.
	%


%

\subsection{{Reviewing Existing DNN Latency Optimizations}}
\label{ss:related_works}


%

Before diving into the technical part, we discuss existing research efforts directly related to our work.
Based on aforementioned analysis, we can divide the current DNN latency optimization works into two major types.


One type of works simplified their latency optimization objective as FLOPs reduction.
	For example, convolutional filter pruning first proposed to remove the non-significant filters to reduce the model parameter volume as well as FLOPs and therefore speed up the system~\cite{lihao}.
	Following that, many works designed different filter significance criteria, such as channel pruning~\cite{prune1}, geometric mean pruning~\cite{prune2}, and feature map rank based pruning~\cite{hrank}.
	However, due to the lack of understanding of intrinsic hardware runtime mechanisms, such FLOPs reduction can only translate to sub-optimal latency performance as mentioned in~\cite{net_adapt}.

The other type of work aimed at the ``end-to-end'' hardware-aware optimization, which treated the DNN execution on GPUs as a black box and leveraged the overall system real-time performance to guide DNN architecture search.
	For example, NetAdapt~\cite{net_adapt} and Partial order pruning~\cite{partial} conducted latency measurement and built look-up tables to estimate the DNN model's general execution latency, which was used to guide DNN optimization with the sampling-based configuration search process.
	To improve the search efficiency, some works like ProxylessNAS~\cite{proxylessnas} and SinglePath~\cite{single_path} designed differentiable latency modeling and integrate it into the optimization objectives.
	Despite different search policies, most methods utilized similar end-to-end profiling-oriented DNN structure configuration, but is limited to those methods' lack of understanding of intrinsic hardware mechanisms, especially during runtime operation.

Beyond conducting end-to-end latency optimization, we dive into the GPU thread-level to analyze the DNN runtime performance and reveal one of the root causes for the sub-optimal latency reduction on existing works~\cite{hrank, effnet}.
	Based on the deep understanding of both algorithm deployment and hardware operation, we further propose a holistic optimization methodology that leads to practical latency reduction.

\begin{figure}[!tb]
	\centering
	\vspace{2mm}
	\includegraphics[width=3.3in]{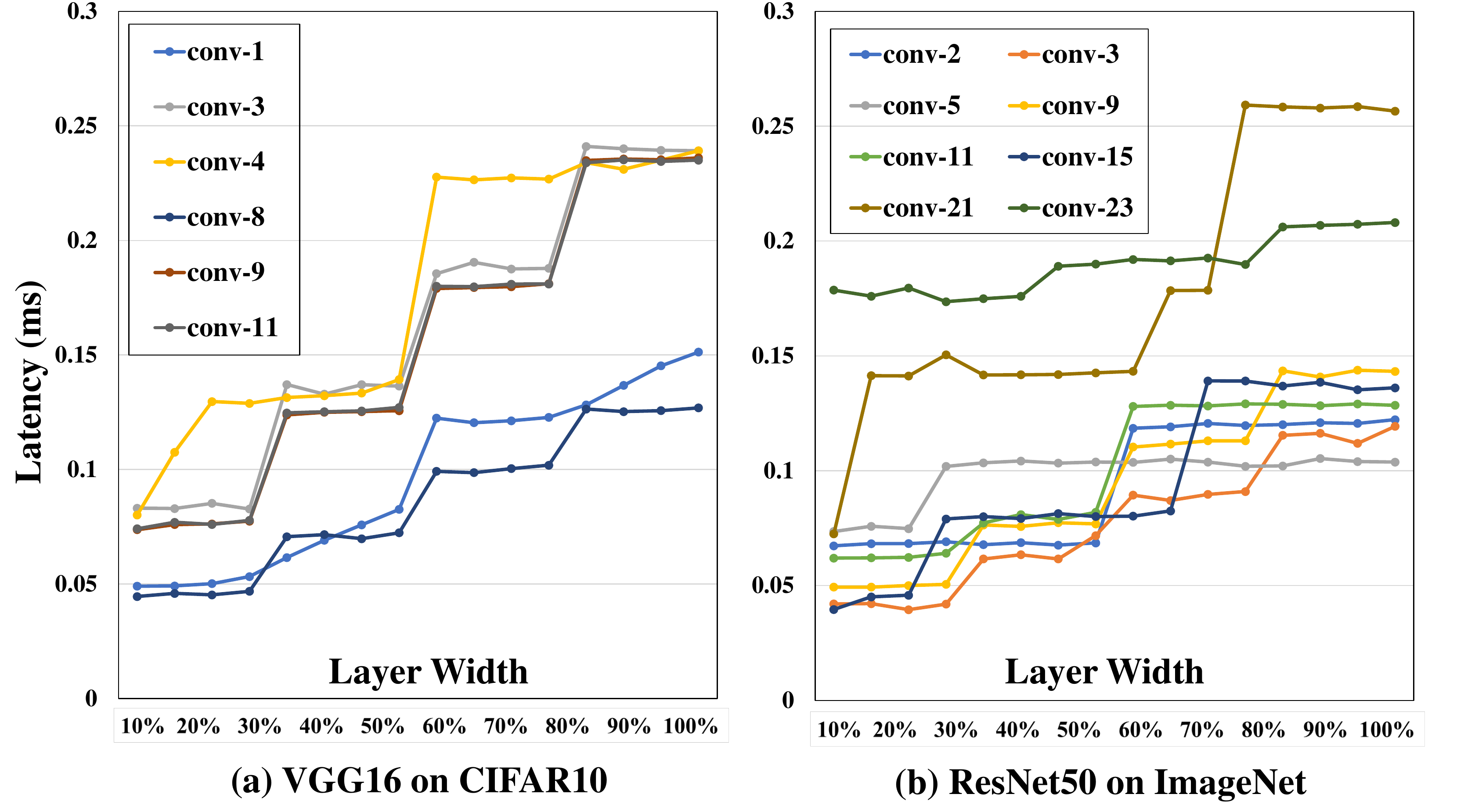}
	\vspace{-5mm}
	\caption{Latency Staircase Profiling Results.}
	\label{fig:case_study}
	\vspace{-3mm}
\end{figure}

\section{{Latency Staircase: GPU Tail Effect on DNNs}}
\label{sec:3}


In this section, we further examine the aforementioned latency staircase phenomena (\S\ref{ss:observation}), and reveal the critical DNN execution mechanism on GPUs, which causes the mismatch between FLOPs and latency reduction (\S\ref{ss:deploeyment}).
	With complementary latency modeling and verification, we also shed light on potential optimization approaches  (\S\ref{ss:verification}).

\begin{figure*}[!t]
	\centering
	\includegraphics[width=6.5in]{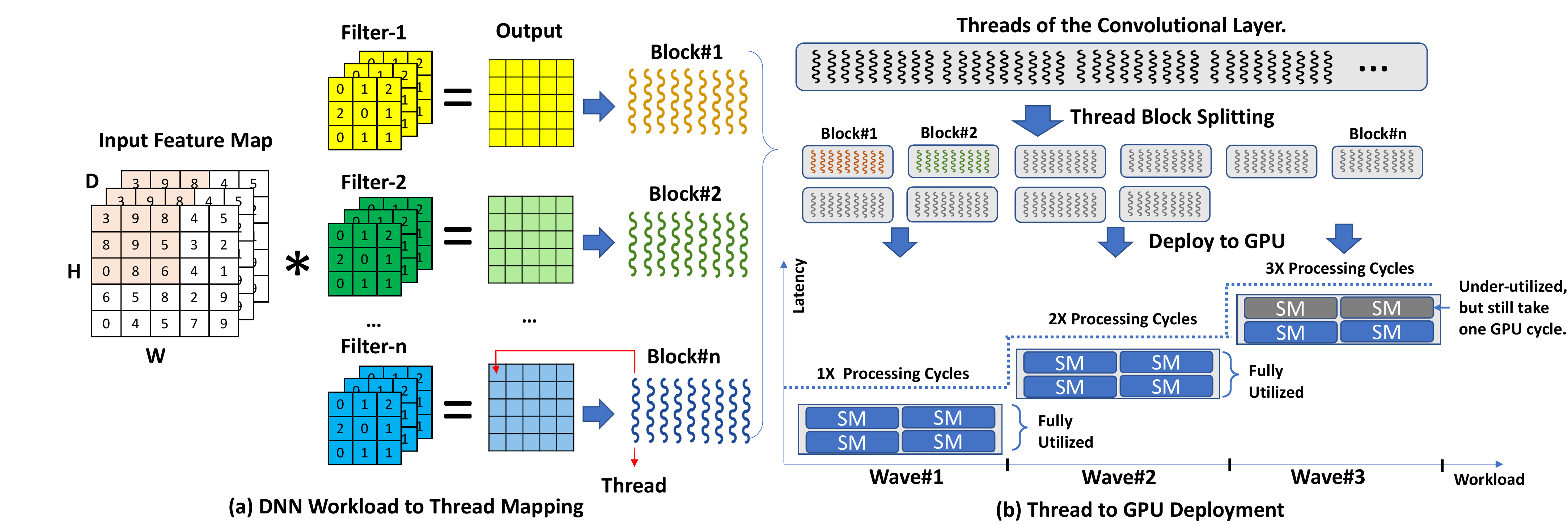}
	\vspace{-1mm}
	\caption{The intrinsic mechanism of DNN deployment on GPUs and potential performance degradation with GPU tail effect.
	(a) In model-to-thread mapping, each convolutional filter is mapped with one thread block. And each thread is responsible for the calculation of one spatial-wise output cell.
	(b) The minimal deployment unit is usually from the thread block to SM\@. 
	For larger DNN models, it will take multiple GPU cycles to finish the workload in waves.
	When the last workload wave cannot fully occupy the GPU SMs, the \textit{GPU tail effect} emerges with the last wave of under-utilized SMs, which will still cost one full GPU processing cycle.
	}
	\label{fig:tailing}
	\vspace{-2mm}
\end{figure*}

\subsection{A Glimpse of Latency Staircase Issue}
\label{ss:observation}

In addition to the experiment results shown in Fig.~\ref{fig:1}, we profile the runtime latency of two representative DNN architectures on an NVIDIA Titan-V GPU: (1) VGG16 on CIFAR10 (32x32x3) and (2) ResNet50 on ImageNet (224x224x3).
	For each DNN architecture, we measure each convolutional layer's runtime latency and repeat the measurement with different layer width settings (obtained by filter pruning).
	Both models are implemented using PyTorch-1.6 with CUDA-10.0 and CuDNN-7.6.5 backend.

\vspace{1mm}
The latency profiling results are shown in Fig.~\ref{fig:case_study}.
%
We find that the latency curves differ from each other significantly \textit{w.r.t} different convolutional layers.
	Nevertheless, all of them show a similar staircase-like growing trend as we gradually increase the layer width.
	Taking the latency curve of the layer of conv-4 on VGG16 as an example, we can see it grows rapidly as we increase the layer width from 10\% to 20\%.
	Through the latency staircase, we term this period as a {\it growing interval}.
	It then enters a {\it steady interval}, where the latency stays still when the layer width grows from 20\% to 50\%.
	The latency curve repeats this stair-case growing pattern as we further increase the layer width to 100\%.

Through these experimental results, we have three major findings:
	$i)$ The latency staircase phenomena widely exist in different DNN architectures' realtime performance;
	$ii)$ The latency performance is highly related to the DNN layer configurations;
	$iii)$ Blindly DNN structure optimization (\textit{e.g.}, reducing layer width) may not always benefit the latency.
	Therefore,  a thorough understanding of the practical DNN deployment on GPUs and the latency staircase phenomena' root cause would effectively guide a comprehensive optimization for the system latency reduction.

\begin{algorithm}[b!]
	\caption{The Convolutional Layer Operation.}
	\label{alg:conv}
	\begin{algorithmic}[1]
		\STATE {\bfseries Input:} Input $I[H,W,D]$, $N$ Filters $F_i[h,w,D]$.
		\STATE {\bfseries Output:} Output $O[H,W,N]$.
		\FOR{Filters $i=1$ {\bfseries to} $N$ ({$\gets$ \textbf{blocks}})}
			\FOR{Input Height $j=1$ {\bfseries to} $H$ ({$\gets$ \textbf{thread.idx}})}
				\FOR{Input Width $k=1$ {\bfseries to} $W$ ({$\gets$ \textbf{thread.idy}})}
					\STATE --- Per Thread {Kernel Func. Starts.}---
					\STATE $O[j,k,i]$ = 0.
					\FOR{Kernel height $x=1$ {\bfseries to} $h$}
						\FOR{Kernel width $y=1$ {\bfseries to} $w$}
							\FOR{Kernel depth $z=1$ {\bfseries to} $D$}
								\STATE $O[j,k,i] \mathrel{+}= I[x,y,z] \times F_i[x,y,z]$
								\STATE --- Per Thread {Kernel Func. Ends.}---
							\ENDFOR
						\ENDFOR
					\ENDFOR
				\ENDFOR
			\ENDFOR
		\ENDFOR
	\end{algorithmic}
\end{algorithm}

\subsection{Understanding DNN Execution on GPU Runtime}
\label{ss:deploeyment}

To understand the root cause of this latency staircase issue, we conduct a ``full-stack'' examination of DNN execution on GPUs.
	Specifically, the examination covers two major execution stages: $i)$ DNN deployment to programming threads, and $ii)$ programming threads to GPU mapping.
	As the latency bottleneck of a DNN lies in convolutional layers~\cite{time_from_conv1}, we mainly focus on the convolutional layer computation in this work\footnote{The illustrated example is direct convolution, the major kernel scheduling implementation for common DNN layers in current CUDA\@. For simplicity, we show the case with batch size=1. Same mechanisms apply for larger batch sizes as we will show later.\vspace{-3mm}}.


\textbf{DNN Deployment with Programming Thread Mapping.}
%
	As illustrated in Algorithm~\ref{alg:conv}, the convolutional layer operation involves $N$ filters convolving with the input feature map $I$, which can be implemented by a multi-level loop calculation.
	The computations shown in the three inner loops (line 8---11) are the atom unit running in the same thread.
	While the computation in the three outer loops (line 3---5) corresponds to different filters or pixels on the feature map. Hence they are parallelized using multiple threads.
	As a rule of thumb, the number of filters ($N$) usually determines the number of thread blocks in need, and the height ($H$) and width ($W$) of the input determine the number of threads in each thread block.
	Fig.~\ref{fig:tailing}(a) illustrates how the convolutional workload is divided and mapped to threads.

\textbf{Programing Thread to GPU Mapping.}
Threads are then mapped to GPUs in the granularity of the thread block.
	As shown in Fig.~\ref{fig:tailing} (b), threads are first grouped into blocks, and then loaded to SMs on the GPU\@.
The mapping can be ``One-to-One'' or ``N-to-One.''
The maximum number of blocks loaded on one SM is determined by its physical capacity (\textit{e.g.}, number of registers, size of share memory).
The DNN model size grows rapidly, which renders the number of thread blocks in need exceed the GPU capacity.
	Therefore the GPU will divide these thread blocks into multiple consecutive {\it waves}, and run these waves in sequence.

\textbf{GPU Tail Effect on DNN.}
In general GPU computing scenarios, the ``GPU tail effect'' happens when the thread blocks on the last wave did not fully occupy all SMs on the GPU.
	However, the last wave's processing-cycle still costs the same time with the full-wave ahead, as the processing cycle depends on the wave that takes the longest running time.

	Through the ``full-stack'' DNN execution mechanism examination, we redefine the GPU tail effect with the DNN model structure configuration as the unstructured DNN execution deployment to GPU operation threads and processors.
	It is worthy to note that,  the GPU tail effect with DNN is increasingly prominent with light-weight DNN design especially in the embedded domain, while embedded GPUs are becoming increasingly powerful with larger computing unit capacity
	Therefore, without dedicated optimization, the mismatch between the algorithm deployment and hardware utilization can be increasingly severe.


\subsection{Latency Staircase Modeling \& Verification}
\label{ss:verification}


By revealing the GPU tail effect with DNN, we can build up a dedicated DNN latency staircase model.
	With the verification of the effectiveness, such a model can effectively guide latency-aware DNN optimization.

\textbf{DNN Latency Staircase Modeling on GPUs}
Based on the above analysis, the GPU runtime latency $L$ of one convolutional layer could then be modeled as follows:
\begin{equation}
\small
\begin{split}
	\vspace{-1mm}
	& L = \Delta L \times \lceil ~B \div S~ \rceil, \\
	\text{where}~ B = & \frac{\text{Number of threads per filter} \times F} {\text{Number of threads per block}}.
	\label{eq:3}
\end{split}
\end{equation}
\normalsize
In Eq.~\ref{eq:3}, $\Delta L$ is the duration of one processing cycle for each SM to finish one thread block, $B$ is the number of thread blocks for this layer, and $S$ is the number of SMs per GPU.
The $\lceil\cdot\rceil$ is the rounding-up function which returns the least integer greater than or equal to the input, e.g., $\lceil~2.1~\rceil$ = 3, denoting the latency ceiling effect caused by the last GPU tail.
For one convolution layer, the overall number of blocks $B$ is determined by two factors: (i) the overall workload, which depends on the number of threads per filter $\times$ the number of filters $F$, and (ii) per block workload, \textit{i.e.}, number of threads per block.
	Therefore, both filter amount and input shape can influence the per-filter workload, translating to a different number of blocks for this layer during deployment.

\begin{figure}[!t]
	\centering
	\includegraphics[width=3.3in]{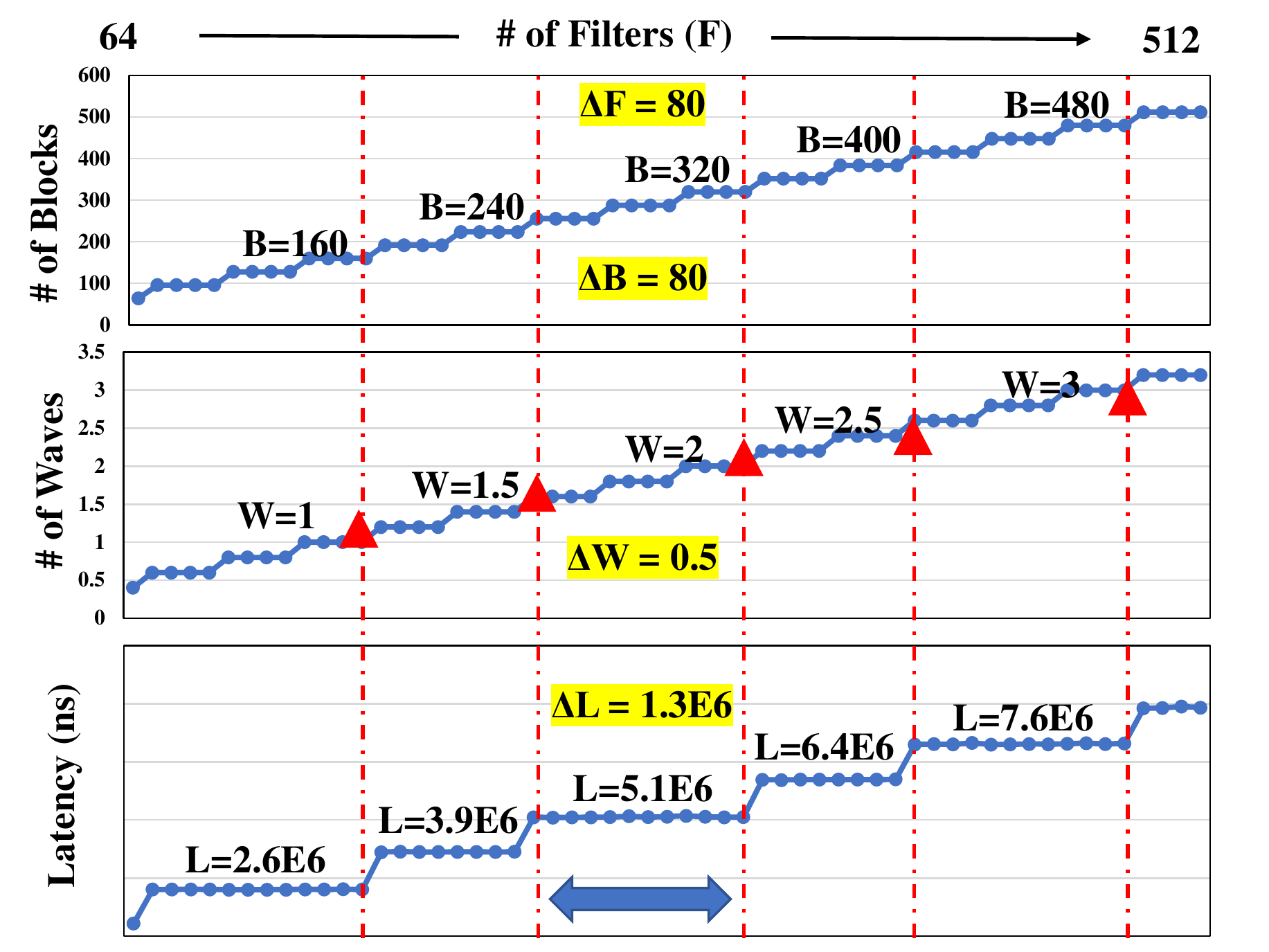}
	\vspace{-5mm}
	\caption{DNN execution performance analysis in perspectives of \textbf{B}: Number of Thread Blocks, \textbf{W}: Number of Waves, and \textbf{L}: Inference Latency.
	Based on the profiling results, we can observe that the runtime latency $L$ increases per $80$ thread blocks, which matches the tested GPU's capacity ($80$ SMs).
	}
	\label{fig:multi_level}
	\vspace{-5mm}
\end{figure}

\textbf{Effectiveness Verification}
To verify the DNN execution latency model established above, we test the model with a set of convolutional layers with different filter numbers from 64 to 512.
We keep the same kernel size (3$\times$3) and the input feature map (64$\times$64$\times$512).
The batch size, by default, is set to 1.
All benchmarks are conducted on a Titan-V GPU\footnote{We will show in later experiments that such phenomenon is general for both high-end and low-end GPUs like Jetson Nano.}.
%
To better understand the results, we also show the number of blocks ($B$) and the number of waves ($W$) corresponding to each layer settings.
The result is shown in Fig.~\ref{fig:multi_level}.
We have the following three key observations.




\vspace{-1mm}
\textit{Verification 1}: The number of blocks ($B$) grows with the number of filters ($F$). For example,  $\Delta F = 80$ leads to $\Delta B = 80$. Therefore, the current layer deployment has mapped its convolution filters one-by-one to the thread blocks, which is similar to Fig.~\ref{fig:tailing}(a).

\vspace{-1mm}
\textit{Verification 2}:  With the number of blocks ($B$) growing, the latency increases with a granularity of 80 blocks $\Delta B = 80$, which is the full GPU capacity (80 SMs). This verifies that the latency increases with the step size of $S$ in Eq.~\ref{eq:3}.

\textit{Verification 3}: For a wave in one flat interval, the latency remains the same regardless of how many blocks are in the wave or how many SMs are actually utilized. This verifies the latency ceiling effect, \textit{i.e.}, the GPU processing cycle per wave depends on the longest time that SM takes.


The above latency performance analysis, mechanism modeling, and effectiveness verification reveals the actual correlation between DNN model structure and GPU runtime operation.
	By filling the algorithm and hardware gap, in the next section, we propose a GPU-aware model optimization methodology to eliminate the GPU tail effect and achieve better latency optimization performance.

\vspace{-2mm}
\section{{Eliminating the GPU Tail: Hardware-aware optimization}}
\label{sec:4}
In this section, the proposed methodology is firstly presented with the design challenges (\S\ref{ss:design_consideration}), followed by the specific design rules (\S\ref{ss:metrics}).
	Finally, the latency-aware DNN optimization methodology is presented (\S\ref{ss:optimization}).




\subsection{Design Challenges}
\label{ss:design_consideration}

When designing the model optimization technique for runtime latency reduction, we should jointly address  two critical challenges:
	$i)$ The design should address the GPU tail effect as it often leads to GPU under-utilization. 
	A incomprehensive FLOPs reduction method may not achieve lower runtime latency. For instance, FLOPs reduction effort applied on the steady interval will not improve the runtime latency (\S\ref{ss:observation}).
	$ii)$  The design should also consider the hardware diversity of GPU configurations. Different GPUs may differ drastically in their capacity (\textit{e.g.}, number of SMs). Hence no one-fit-all DNN model configuration exists even for the same model running on different GPU platforms.
	




\begin{figure}[!t]
	\centering
	\includegraphics[width=3.3in]{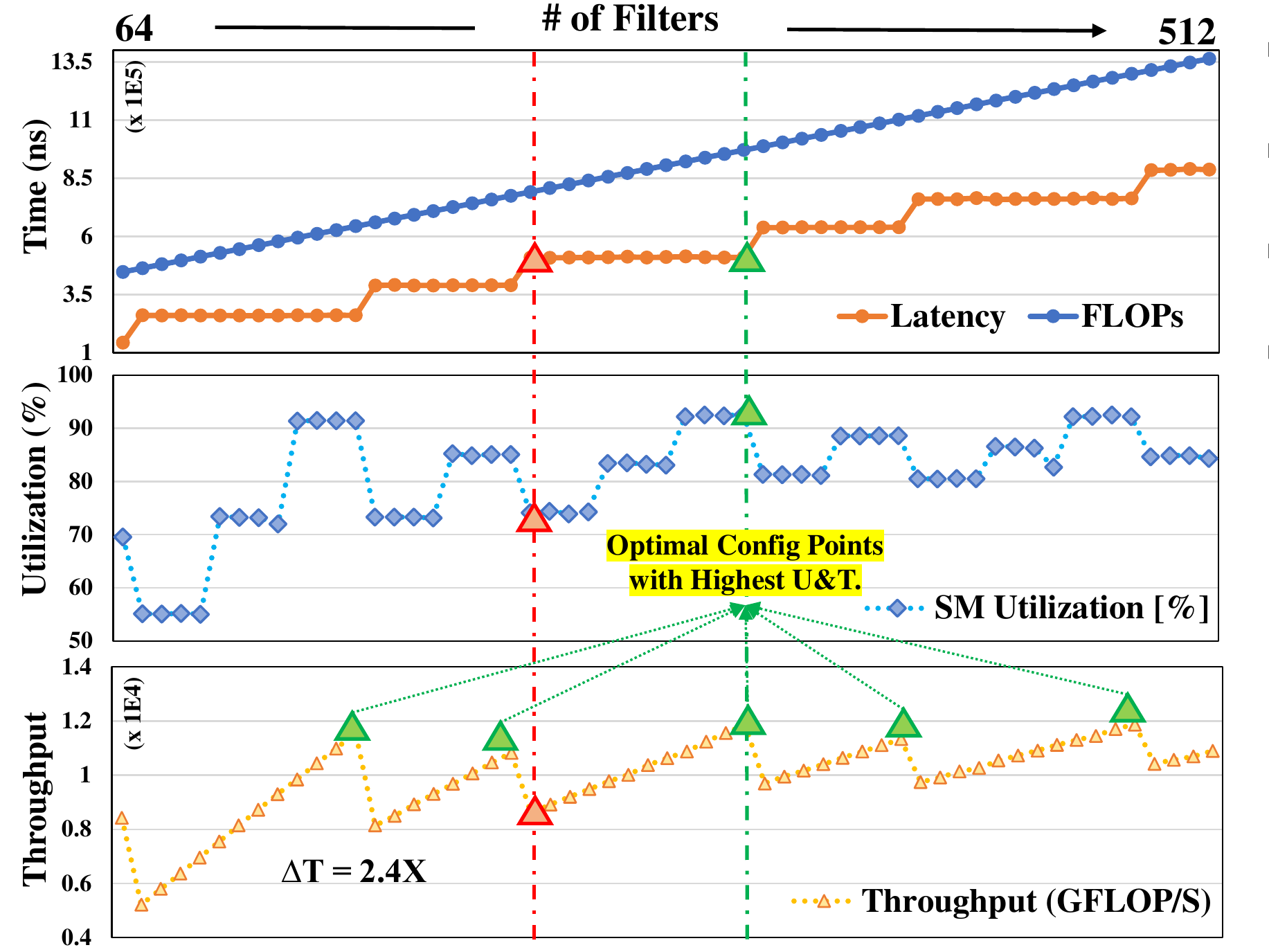}
	\vspace{-5mm}
	\caption{Underlying the latency staircase, the GPU's runtime efficiency changes dramatically, which can be reflected by the SM Utilization (U) and Throughput (T). By being aware of such GPU runtime efficiency guidelines, we can identify the optimal configuration candidates for optimization.}
	\label{fig:multi_level_2}
	\vspace{-2mm}
\end{figure}

\subsection{Combining SM Utilization and GPU Throughput}
\label{ss:metrics}


\textbf{Combined Design} The GPU tail effect leads to an inefficient utilization of SMs on GPU.
Hence an intuitive idea is to employ the SM utilization rate to guide the DNN model optimization, so that we can eliminate the tail effect and achieve the optimal latency reduction.
In practice, however, the profiled SM utilization can often be imprecise\footnote{For example, due to the non-divergence execution property of GPUs, certain SMs should have been idle may still run non-useful instructions. The current NVIDIA profiling toolset can take this as the utilized status and can report imprecise utilization statistics.}, hence cannot serve as a reliable cue to guide the DNN model structure optimization.

Therefore, we leverage both SM utilization and GPU throughput to guide the model optimization. 
    The effectiveness of using both SM utilization and GPU throughput can be better illustrated by  Fig.~\ref{fig:multi_level_2}:
    Taking the middle latency staircase as an example, the left and right edge points denote two layer configurations within the same wave, but with the most under-utilized GPU tail (left) and no GPU tail (right). 
	As a result, the left point shows the lowest GPU utilization (73\%) and throughput (8.5 TFLOP/s).
	While the right point shows the highest SM utilization  (93\%), and near upper-bound GPU throughput (11.8 TFLOP/s).
Thus, by jointly considering the configurations with the maximum utilization and throughput, we can effectively identify the optimal layer configuration without any tail effect.

\textbf{Design Formulation} The optimal configuration for layer $i$ can be obtained by maximizing the utilization $\&$ throughput:

\vspace{-3mm}
\begin{equation}
C_i[m] = \text{Arg} ~\text{max}_m(U_i \times T_i),
\label{eq:candidates}
\vspace{-1mm}
\end{equation}
where $U_i$ and $T_i$ denote the utilization and throughput information for this layer, and $m$ is the number of optimal candidates. 
For example, in Fig.~\ref{fig:multi_level_2}, by considering the maximum $U$ and $T$, we can identify five layers' width configurations that yield the optimal GPU runtime efficiency.

Based on such GPU efficiency guidelines capable of identifying the optimal layer configuration candidates, we then propose our GPU-aware model optimization algorithm.

\subsection{GPU-Aware Model Configuration Optimization}
\label{ss:optimization}


\textbf{Accuracy-Latency Optimization Objectives} We first define the model-level accuracy-latency trade-off objectives.

Naturally, the accuracy latency trade-off can be defined as trading off the \textit{latency gain (LG)} and the accuracy loss.
Also, we use the \textit{parameter gain (PG)} to describe the model capacity expansion and therefore as an accuracy gain approximation.
The latency gain and parameter gain can be estimated based on latency profiling $L_i$ and layer width $R_i$:
\begin{subequations}
\begin{equation}
LG_i = L_i[R_{i, old}] - L_i[R_{i,new}],
\label{eq:lg}
\vspace{-1mm}
\end{equation}
\begin{equation}
PG_i = R_{i, old} - R_{i, ~new},
\label{eq:pg}
\end{equation}
\end{subequations}
where $R_{i,new}$ and $R_{i,new}$ denote the optimized and the original layer configuration, $LG_i$ indicates the latency gain, and $PG_i$ indicates the parameter gain which can be negative if we scale down the layer width.

\begin{algorithm}[!tb]
   \caption{GPU-Aware Model Optimization Algorithm.}
   \label{alg:2}
\begin{algorithmic}[1]
   \STATE {\bfseries Input:} Model's initial layer width configs $r[l]$, latency $L[l][n]$, utilization $U[l][n]$, and throughput $T[l][n]$.
   \STATE {\bfseries Output:} Optimized model configs $R_{new}[l]$.
   \STATE Identify candidates $C_l[m]$ for each layer $l$ by Eq.~\ref{eq:candidates}.
   \STATE Initialize latency \& parameter gain list $LG[l]$, $PG[l]$.
	\FOR{layers $i=1$ {\bfseries to} $l$}
	   \STATE Get $LG_i$, $PG_i$ estimation by Eq.~\ref{eq:lg} and~\ref{eq:pg}.
	   \ENDFOR
	  \STATE Sort the layer index list by $LG[l]$ or $PG[l]$.
	\WHILE {layer index list is not empty}
	   \STATE Pop out layer j with $\text{Arg}\text{max}_j ~LG[l]$.
	   \STATE Scale down $R_{j,new}$ by Eq.~\ref{eq:scale_down} for max latency gain.
	   \WHILE {$\Sigma^l PG(R_{new}) \notin (-\tau, \tau)$}
	   		\STATE Pop out layer k with $\text{Arg}\text{min}_k ~LG[l]$.
	   		\STATE Scale up $R_{k,new}$ by Eq.~\ref{eq:scale_up} to balance param gain.
	   \ENDWHILE
	\ENDWHILE
	  \STATE Get runtime latency evaluation $L_{new}$ of config $R_{new}$.
	  \IF {$L_{new} \leq L_{old} \times \delta$ [$\gets$ Targeted reduction ratio]}
	  \STATE Train and evaluate the model accuracy.
	  \ELSE
	  \STATE Set $\tau ~\text{*=}~ 2$ and repeat the algo. from line 9.
	  \ENDIF
	  \STATE \textbf{Return} Optimized config $R_{new}$.
\end{algorithmic}
\end{algorithm}

\underline{\textit{Accuracy-Oriented Optimization}}: The accuracy-oriented optimization aims to boost the accuracy without incurring any latency overhead. As a larger parameter gain demonstrates better accuracy, the goal can be represented by:
\begin{equation}
\text{Maximize} ~~\sum\nolimits_i^l PG_i, ~~~\text{s.t.}~~~ \sum\nolimits_i^l LG_i \geq 0.
\vspace{-1mm}
\end{equation}

\underline{\textit{Latency-Oriented Optimization}}: For latency-oriented optimization, we aim to reduce the latency while maintaining no or negligible accuracy drop.
The objective can be formulated as maximizing the latency gain while maintaining parameter gain in a tolerable range ($\tau$):
\begin{equation}
\text{Maximize} ~~\sum\nolimits_i^l LG_i, ~~~\text{s.t.}~~~ \sum\nolimits_i^l PG_i \in (-\tau, \tau).
\vspace{-1mm}
\end{equation}


\textbf{Latency-Aware DNN Model Optimization} Based on the previous objective, we then propose our GPU-aware model optimization method. 
We take the latency optimization algorithm to illustrate as following.
The overview is shown in Algorithm~\ref{alg:2}, which includes four major steps
\footnote{The accuracy-oriented optimization follows the similar procedure, which could be found in supp. material.}.

\vspace{-1mm}
\underline{\textit{Step 1. DNN Structure Modification Pre-Analysis:}} Given the DNN model structure, we first profile the full-spectrum guideline metrics (i.e., latency, utilization) for each layer. 
The throughput information can then be derived by theoretical FLOPs and profiled latency.
	The optimal candidate configurations $C_i[m]$ for layer $i$ can be identified by Eq.~\ref{eq:candidates}.
As we show before, the optimal configurations per layer usually consist of only several discrete settings. Thus, we can greatly reduce the search space.

\vspace{-1mm}
\underline{\textit{Step 2. Layer-Level DNN Structure Adjustment:}} For each convolutional layer in the model, the layer width adjustment follows two ways, \textit{i.e.}, either scaling down for latency gain ($LG$) or scaling up for parameter gain ($PG$):
\begin{subequations}
\begin{equation}
R_{i,new\downarrow} = \max(^*C_i[m] < R_{i,old}),
\label{eq:scale_down}
\vspace{-2mm}
\end{equation}
\begin{equation}
R_{i,new\uparrow} = \min(^*C_i[m] > R_{i,old}),
\label{eq:scale_up}
\end{equation}
\end{subequations}
where $R_{i,new}$ and $r_{i,old}$ indicate adjusted and original layer width for layer $i$, and $C_i[m]$ is the optimal candidate list.
The first strategy Eq.~\ref{eq:scale_down} denotes we scale down the layer width to the left optimal configuration candidates. 
By doing so, we can remove the last GPU tail, thus reducing the latency. 
By contrast, the second strategy Eq.~\ref{eq:scale_up} indicates that we can scale up the layer width to fully-occupy the last GPU tail, thus potentially enhancing the model accuracy without incurring any latency overhead.

\vspace{-1mm}
\underline{\textit{Step 3. Model-Level DNN Structure Adjustment:}} One naive solution of maximizing the latency optimization is to scale down all layers according to Eq.~\ref{eq:scale_down}.
	However, this can incur larger negative parameter gain and thus hurt the model accuracy.
	To maintain the parameter gain in the small range $(\tau, \tau)$, we propose a balanced model adjustment strategy by simultaneously scaling down and scaling up different layers in a balanced manner.

Specifically, we maintain two queues of layer indexes ranked by $LG_i$ and $PG_i$. 
	To maximize $LG$ while meeting the $PG$ constraints, we greedily pop up layers with top $LG_i$ to scale down by Eq.~\ref{eq:scale_down}, while simultaneously pop up layers with the least $PG_i$ to scale up by Eq.~\ref{eq:scale_up} to balance the negative parameter gain.
	The model optimization will stop when all layers are popped out and adjusted accordingly.

\underline{\textit{Step 4. DNN Structure Determination:}}  After generating the new model configuration, we first conduct a runtime latency check to filter out configurations that cannot meet the model latency reduction requirements $L_{new} \leq L_{old} \times \delta$.
	The $\delta$ here denotes the targeted latency reduction ratio.
	If the current model cannot meet the latency requirements, the algorithm can repeat by loosing the constraints, e.g., allowing larger parameter gain tolerance $\tau$.
	It allows for more aggressive optimization to meet the latency reduction target.

\begin{figure}[!tb]
	\centering
	\includegraphics[width=3.3in]{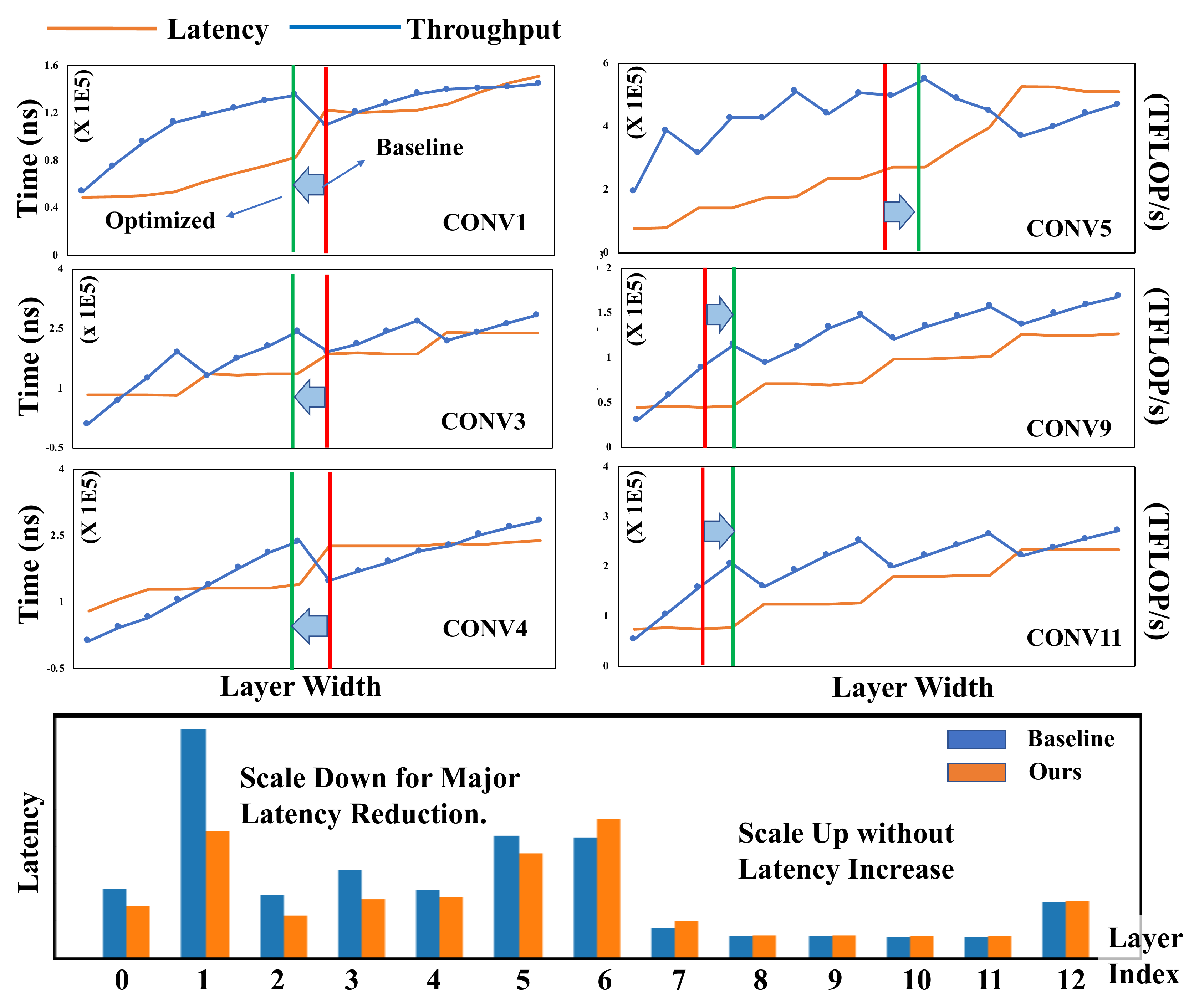}
	\vspace{-8mm}
	\caption{VGG pruning configuration optimization on HRank. We slightly scale down Conv1-Conv4 using minimal accuracy drop to trade for major latency reduction, while scale up Conv5-Conv12 to compensate the accuracy drop without obvious latency increment.}
	\vspace{-5mm}
	\label{fig:vgg_pruning}
\end{figure}

\textbf{Optimization Demonstration with A Case Study} To illustrate the algorithm, we demonstrate one of the SOTA method's~\cite{hrank} sub-optimal optimization results on GPUs and then show our optimization mechanisms.

Fig.~\ref{fig:vgg_pruning} shows the per-layer configuration from \cite{hrank} (red) and our optimized ones (green). 
	As we can see, for CONV1, 3, 4, the original layer width is obviously non-optimal, which leads to the tail effect with low GPU throughput. 
	To optimize that, we scale down the layer width slightly to meet the left optimal configuration and thus eliminate the GPU tail.
	By such adjustment, the latency of these layers shows a dramatic reduction.
	Then as the compensation, we scale up certain later layers, \textit{e.g.}, CONV5, 9, 11, to balance the negative parameter gain without incurring obvious latency overhead.
As a result of such configuration optimization, we can reach the similar accuracy (92.9\%) as the baseline (93.1\%) but deliver \textbf{17.7\%} overall latency reduction, which demonstrates better accuracy-latency trade-offs than the baseline method.

\subsection{Applicability in Different Optimization Methods}

As an orthogonal hardware-oriented optimization dimension, our configuration optimization method can be applied to enhance existing model optimization methods:

\textbf{Advancing Filter Pruning} Filter pruning methods usually design the filter pruning criteria based on the accuracy influence, forming usually a continuous search space for each layer's width (i.e., how many filters to prune).
    Our method can be integrated into such methods by providing per layer optimal candidates list, thus providing a discrete pruning space. 
    The further training strategies can remain the same.

\textbf{Advancing NAS} For NAS task, on one hand, our method can provide similar discrete search space for layer configurations, thus enhancing both the search efficiency and the final model's GPU-utilization performance. 
    On the other hand, we can also conduct post-optimization by further adjusting the existed NAS model architectures to achieve better latency-accuracy trade-off.

%





\section{{Experimental Evaluation}}
\label{sec:exp}
\vspace{-2mm}

\subsection{Experimental Setup}

We conduct experiments using the software stack including PyTorch-1.6, CUDA-10.1 and CuDNN-7.6.5.
	For GPU generality, we evaluate three GPUs from high-end (Titan-V, P6000) to embedded ones (Jetson Nano). 
	The specification comparison of them is shown in Table~\ref{tab:gpus}.
	
We apply our method into two common model optimization methods: Filter pruning and NAS. 
	For the pruning task, two state-of-the-art pruning methods are chosen as baselines, including HRank~\cite{hrank} and SoftPruning~\cite{soft}.
	For the NAS task, we mainly apply our model optimization method onto the EfficientNet series~\cite{effnet}, one of the SOTA efficient model architectures.

\subsection{Evaluation in the Pruning Task}
We compare our method on filter pruning tasks with HRank~\cite{hrank} and SOFT~\cite{soft}.


\vspace{-1mm}
\underline{\textit{Latency Reduction}}: As Table~\ref{table:latency} presents, by optimizing the pruning configurations, our method achieves consistently lower latency than baseline methods (\textbf{11.3\%} to \textbf{17.7\%} latency reduction) with negligible accuracy influence.  
Our latency improvements can also be explained from the following two aspects.

\vspace{-1mm}
\underline{\textit{Throughput Maximization}}: 
    As Table~\ref{table:latency} shows, our method shows consistently higher GPU throughput than the baselines (see (\textit{FLOP/s}) column). Take VGG16 architecture as an example, the optimized model using our method can achieve \textbf{3.90 TFLOPs} on GPU, showing \textbf{1.6$\times$} throughput increase compared to the baseline method, 2.41 TFLOPs. 

\vspace{-1mm}
\underline{\textit{Balanced Parameter Gain (PG)}}: 
    In addition to lower latency, our method can even maintain more parameters and improve the baseline accuracy. 
    As (\textit{Params}) column shows, our optimized models have maintained the parameter gain either in positive range or small negative range, which ensures our models' accuracy performance. 
    

\begin{table}[]
\centering
\renewcommand\arraystretch{1.0}
\caption{The Evaluated GPUs and Specifications.}
\vspace{1mm}
\begin{adjustbox}{width=0.48\textwidth}
\begin{tabular}{ccccc}
\toprule
GPU Name & Archicture & \#SMs & \#Cores & Peak FLOP/s \\ \midrule
Titan-V & Volta & 80 & 5120 & 14.9T \\ \hline 
Quadro P6000 & Pascal & 30 & 3840 & 12.0T \\ \hline
Jetson Nano & Maxwell & - & 128 & 0.24T \\ \bottomrule
\end{tabular}
\end{adjustbox}
\label{tab:gpus}
\end{table}

\begin{table}[!tb]
\centering
\vspace{-3mm}
\renewcommand\arraystretch{1.3}
\caption{Latency Optimization on SOTA Pruning Works.
HRank: High Rank Pruning [CVPR20]. SOFT: Soft Pruning [IJCAI18].}
\vspace{1mm}
\begin{adjustbox}{width=0.48\textwidth}
\begin{tabular}{ccccccc}
\toprule
{[}Titan-V{]} & Method & Params & \#FLOPs         & FLOP/s         & Acc.\%        & Time (ns)                 \\ \midrule
\multirow{2}{*}{\begin{tabular}[c]{@{}c@{}}VGG16\\ (CIFAR10)\end{tabular}}    & HRank & 1.90M & 67.0M & 2.41T & \textbf{93.1} & 2.50E6 \\ \cline{2-7} 
              & Ours   & 2.85M  & \textbf{104.1M} & \textbf{3.90T} & 92.9          & \textbf{2.05E6 (-17.7\%)} \\ \midrule
\multirow{3}{*}{\begin{tabular}[c]{@{}c@{}}ResNet56\\ (CIFAR10)\end{tabular}} & HRank & 0.48M & 65.9M & 0.83T & 93.6          & 4.20E6 \\ \cline{2-7} 
              & Ours-1   & 0.50M  & \textbf{79.1M}  & \textbf{1.00T} & \textbf{93.8} & \textbf{3.72E6 (-11.3\%)} \\ \cline{2-7} 
              & Ours-2   & 0.50M  & \textbf{75.1M}  & \textbf{0.95T} & 93.5          & \textbf{3.51E6 (-16.3\%)} \\ \midrule
\multirow{4}{*}{\begin{tabular}[c]{@{}c@{}}ResNet56\\ (CIFAR10)\end{tabular}} & SOFT-1  & 0.53M & 68.8M & 0.88T & 93.1          & 4.08E6 \\ \cline{2-7} 
              & Ours   & 0.43M  & \textbf{71.4M}  & \textbf{0.91T} & \textbf{93.2} & \textbf{3.52E6 (-13.7\%)} \\ \cline{2-7} 
              & SOFT-2   & 0.45M  & 53.1M           & 0.68T          & 92.3          & 3.64E6                    \\ \cline{2-7} 
              & Ours   & 0.43M  & \textbf{66.0M}  & \textbf{0.79T} & \textbf{92.3} & \textbf{3.01E6 (-17.3\%)} \\ \bottomrule
\end{tabular}
\end{adjustbox}
  \scriptsize
  {\raggedright *Note that, SOFT-1 and -2 denotes two configs with different pruning rates from the original paper. The latency is evaluated on Titan-V with batch size = 128.\par}
\vspace{-3mm}
\label{table:latency}
\end{table}


\subsection{Evaluation in the NAS Task}

\begin{table}[!tb]
\centering
\renewcommand\arraystretch{1.3}
\caption{Accuracy Optimization on EfficientNets.}
\vspace{1mm}
\begin{adjustbox}{width=0.48\textwidth}
\begin{tabular}{ccccccc}
\toprule
\multirow{2}{*}{} & \multirow{2}{*}{Method} & \multirow{2}{*}{Acc.\%} & \multicolumn{2}{c}{GPU:Titan-V} & \multicolumn{2}{c}{GPU:P6000} \\ \cline{4-7} 
 &  &  & Time(ms) & FLOP/s & Time(ms) & FLOP/s \\ \midrule
\multirow{2}{*}{\begin{tabular}[c]{@{}c@{}}EfficientNet\\ (ImageNet)\end{tabular}} & B0 & 77.52 & 12.6 & 61.9G & 13.8 & 56.5G \\ \cline{2-7} 
 & Ours & \textbf{81.49 (+3.97)} & 12.7 & \textbf{414.2G} & 14.1 & \textbf{373.0G} \\ \hline
\multirow{2}{*}{\begin{tabular}[c]{@{}c@{}}EfficientNet\\ (ImageNet)\end{tabular}} & B1 & 79.38 & 17.8 & 78.7G & 19.8 & 70.7G \\ \cline{2-7} 
 & Ours & \textbf{82.08 (+2.7)} & 18.0 & \textbf{442.7G} & 19.9 & \textbf{396.0G} \\ \hline
\multirow{2}{*}{\begin{tabular}[c]{@{}c@{}}EfficientNet\\ (ImageNet)\end{tabular}} & B2 & 80.18 & 18.2 & 109.9G & 19.9 & 100.5G \\ \cline{2-7} 
 & Ours & \textbf{82.32 (+2.14)} & 18.4 & \textbf{547.3G} & 20.4 & \textbf{488.2G} \\ \bottomrule
\end{tabular}
\end{adjustbox}
\label{table:effnet}
\end{table}

In this part, we apply our optimization method to further optimize the NAS network's performance on GPUs.  
	Specifically, we optimize the EfficientNet series of model structures to achieve better accuracy latency trade-offs.
	The evaluation results are shown in Table~\ref{table:effnet}.
	The latency evaluation is conducted on both Titan-V and P6000 GPUs with batch size = 1 to simulate the real-time performance.

\underline{\textit{Accuracy Maximization}}: As Table~\ref{table:effnet} shows, with the similar runtime latency, our optimized EfficientNet model could achieve much better accuracy (\textbf{+3.97\%}, \textbf{+2.7\%}, \textbf{+2.14\%}, respectively) on the challenging ImageNet dataset. 

\begin{figure}[!tb]
	\centering
	\includegraphics[width=3.1in]{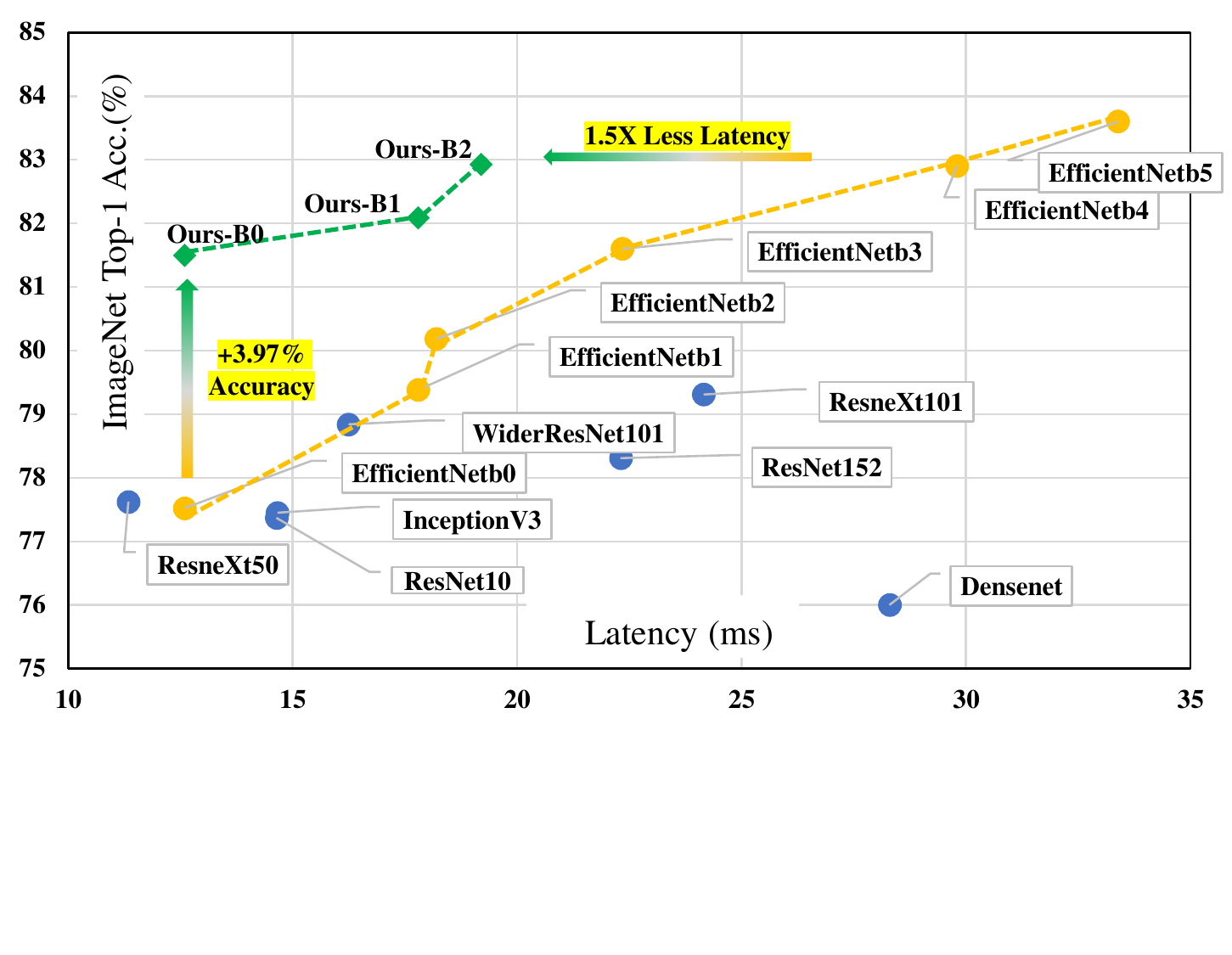}
	\vspace{-18mm}
	\caption{ImageNet Accuracy Latency Trade-Off Comparisons. Compared to baseline EfficientNets (yellow), our GPU-aware optimized models achieve better accuracy latency trade-offs. Evaluated Platform: NVIDIA Titan-V GPU.}
	\vspace{-3mm}
	\label{fig:effnets}
\end{figure}

\underline{\textit{Better Latency Accuracy Trade-offs}}: We composedly compare the model accuracy-latency performance with different model structures, as shown in Fig.~\ref{fig:effnets}.
	From the \textit{accuracy} perspective, our optimized B0 to B3 models achieve the highest accuracies than most baselines under the same latency constraints.
	From the \textit{latency} perspective, our models' latency is 1.5x less than the models which achieve the same accuracy level, \textit{e.g.}, EfficientNet-B3 and B4.
	Therefore, both perspectives demonstrate the effectiveness of our method in achieving better accuracy latency trade-offs on GPUs.

\textbf{Optimization Explanation}
Here we explain our optimization mechanisms for the EfficientNet series of networks.
	When conducting inference on GPUs, we observe that EfficientNet models are highly under-utilizing the GPUs with low throughputs, \textit{e.g.}, only 61.9 - 109.9 GFLOP/s as shown in Table~\ref{table:effnet}. 
	Such under-utilization GPU capacity enables us to increase the network capacity for better model accuracy but without incurring latency overheads.

Specifically, EfficientNet series of networks have three scaling up dimensions: layer width ($w$), input resolution ($r$) and network depth ($d$)~\cite{effnet}. 
	Out of the three dimensions, we scale up both ($w$) and ($r$) to increase the network capacity. 
	The depth ($d$) dimension, however, is kept same with the baseline model, as network layers are run sequentially and depth scaling would inevitably incur latency increase.
	By such width and input resolution scaling, the optimized models could have much larger capacity and improved GPU utilization and throughput, \textit{e.g.}, 414.2 - 547.3 GFLOP/s, but with the similar latency as the baseline model, as shown in Table~\ref{table:effnet}.

\subsection{Generalizability across DNN Architectures}

In this part, we conduct generalizability evaluation of DNN latency staircase in terms of  DNN hyper-parameters, including batch sizes, input resolutions and filter shapes. 

\begin{figure}[!tb]
	\centering
	\includegraphics[width=3.0in]{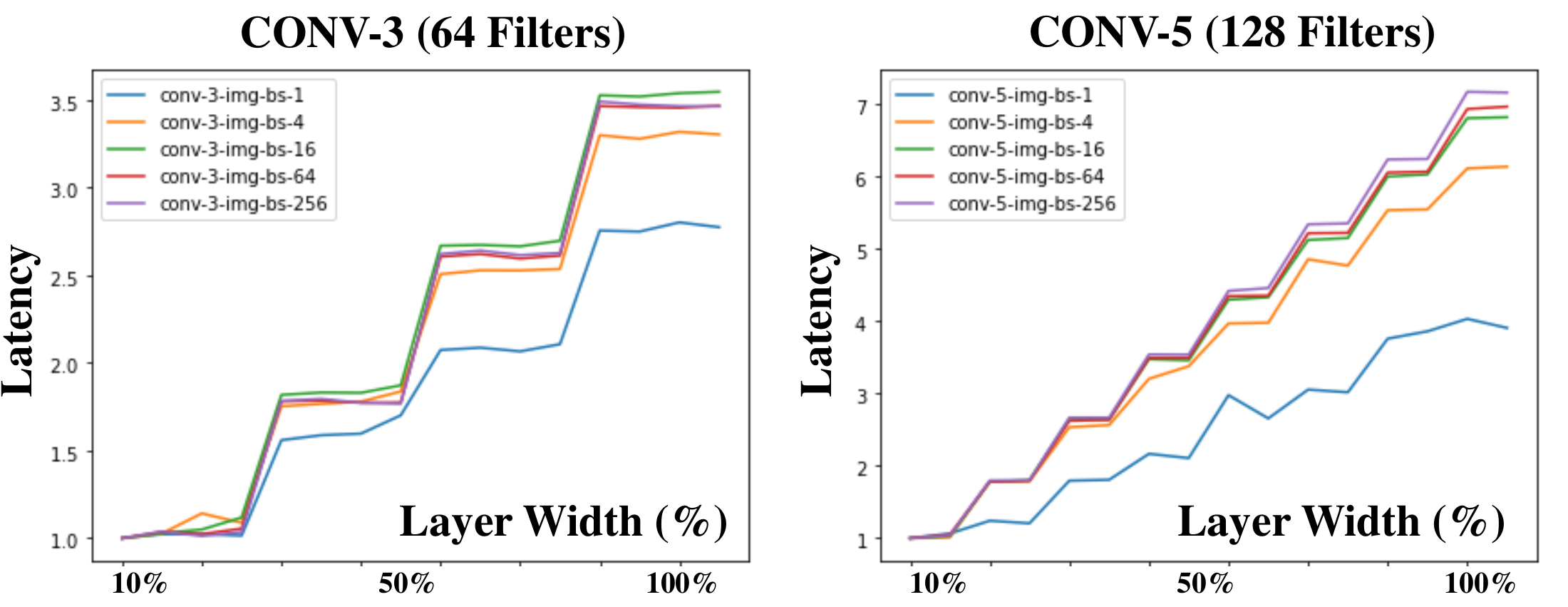}
	\vspace{-3mm}
	\caption{Latency Staircase Generalizability w.r.t \textbf{Batch Size} (1 to 256 in sub-figure) and \textbf{Number of Filters} (CONV3 vs. CONV5).
	}
	\label{fig:batch_size}
	\vspace{-2mm}
\end{figure}

\textbf{Batch Size Variation}
%
	Fig.~\ref{fig:batch_size} shows the latency \textit{w.r.t.} varied batch size from 1 to 256 fro two CONV layers of VGG16 on ImageNet resolution. The latency staircase consistently exists for all layers with different batch sizes. Specifically, layers with larger batch size 
	incurs a longer processing cycle for a wave, leading to a higher latency staircase. 

\textbf{Number of Filters Variation} 
Fig.~\ref{fig:batch_size} also shows that CONV-5 has more levels of staircase compared to CONV-3. The reason is that CONV-5 has more filters (128) than CON (64). Since the number of filters determines the number of thread blocks, 
more sequential GPU waves are needed with a larger number of filters, leading to more levels of staircase.

\begin{figure}[!tb]
	\centering
	\includegraphics[width=3.0in]{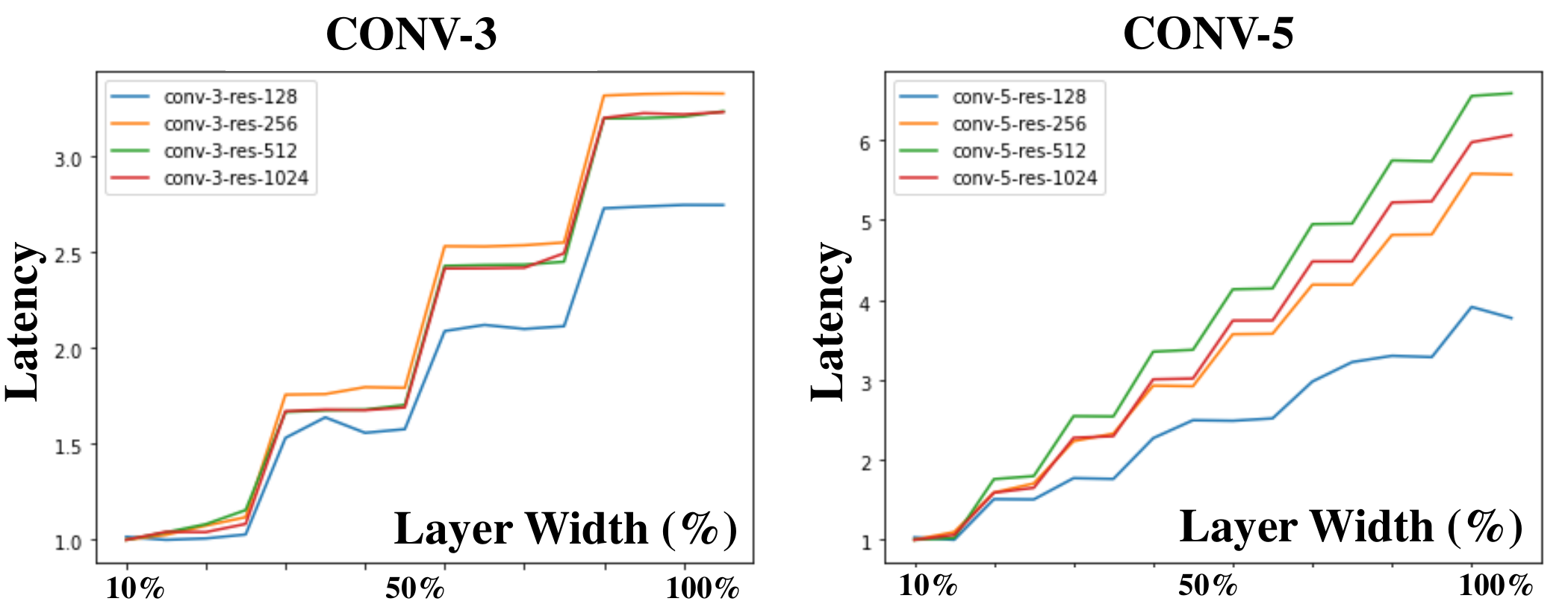}
	\vspace{-3mm}
	\caption{Latency Staircase Generalizability w.r.t \textbf{Resolution}.}
	\label{fig:resolution}
	\vspace{-2mm}
\end{figure}

\textbf{Input Resolution Variation}
Similar latency evaluation is also conducted w.r.t varied input resolution from 128 to 1024 for CONV-3 and CONV-5 in Fig.~\ref{fig:resolution}. The latency staircase still consistently shows in both low to high resolutions.

\begin{figure}[!tb]
	\centering
	\includegraphics[width=2.5in]{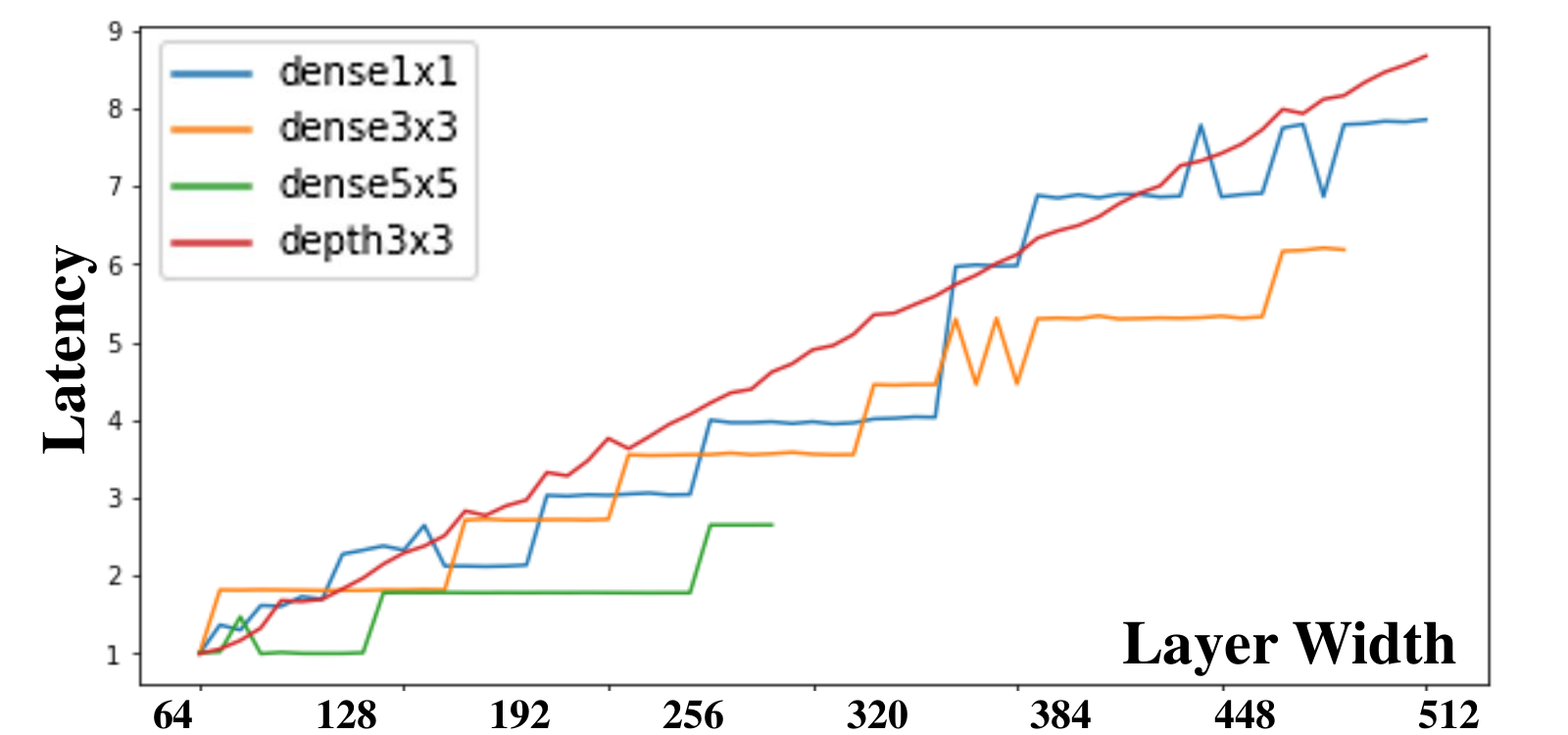}
	\vspace{-3mm}
	\caption{Latency Staircase Generalizability \textit{w.r.t} \textbf{Filter Shape}.}
	\label{fig:shape}
	\vspace{-4mm}
\end{figure}

\textbf{Filter Shape Variation}
In addition, we vary the filter shapes from 1x1 dense filter to 3x3 dense filter, 5x5 dense filter and 3x3 depthwise filter.
    The difference between dense and depthwise filter is the dense kernel is of shape $n \times n \times 512$, while the depth-wise kernel is only $n \times n \times 1$.
Fig.~\ref{fig:shape} shows the generalizability of latency staircase w.r.t. 1x1, 3x3 and 5x5 dense filters.
One exception here is the depthwise 3x3 filter, whose latency results demonstrate a relatively linear pattern.
	The reason is that the depthwise filter is more light-weight compared to dense kernels.
	Therefore, the GPU processing cycle for depthwise workload per wave is relatively small, leading to no dramatic latency increment, i.e., staircase.
	However, as depthwise filters are usually combined with 1x1 dense filters, e.g. in MobileNets, MnasNets, etc~\cite{mobnet, mnasnet}, the latency staircase still exists for such models.

\vspace{-2mm}
\subsection{\textbf{Generalizability across GPU Platforms}}

We evaluate the generalizability of our methods on both high-end GPU, P6000 and Embedded GPU, Jetson Nano. Their specifications in Table~\ref{tab:gpus}.
Although Jetson Nano has much less capacity with only one unified chip (similar to one SM), and much less CUDA cores (only 128 CUDA cores), our method can be generally applied and help achieve better accuracy-latency trade-offs on both platforms. 
	The results are shown in Table~\ref{table:p6000} and Table~\ref{table:jetson}.

As the results show, we could achieve \textbf{9.0\%} to \textbf{27.2\%} latency reduction than the baseline methods, while maintaining similar accuracy for VGG16 and ResNet56 on the P6000 GPU.
	Similar latency benefits (\textbf{13.3\%} to \textbf{20.5\%}) could be achieved for the Jetson Nano GPU, demonstrating the generality of our method across different GPU platforms.

The above results demonstrate the effectiveness and generalizability of our GPU efficiency guideline and algorithm design.
	Without any GPU-specific assumptions, our methods could be applied to a spectrum of GPUs to enhance the current DNNs' accuracy-latency trade-off.

\begin{table}[!tb]
\centering
\renewcommand\arraystretch{1.2}
\caption{Generalizability Evaluation on the Pascal P6000 GPU.}
\vspace{1mm}
\begin{adjustbox}{width=0.48\textwidth}
\begin{tabular}{cccccc}
\toprule
{[}P6000{]}                                                                   & Method & Params & FLOP/s         & Acc.\%        & Time (ns)                 \\ \midrule
\multirow{2}{*}{\begin{tabular}[c]{@{}c@{}}VGG16\\ (CIFAR10)\end{tabular}}    & HRank  & 1.90M  & 1.68T          & \textbf{93.1} & 4.15E6                    \\ \cline{2-6} 
                                                                              & Ours   & 2.04M  & \textbf{1.86T} & 92.9          & \textbf{3.02E6 (-27.2\%)} \\ \midrule
\multirow{2}{*}{\begin{tabular}[c]{@{}c@{}}ResNet56\\ (CIFAR10)\end{tabular}} & HRank  & 0.48M  & 0.84T          & 93.6          & 5.84E6                    \\ \cline{2-6} 
                                                                              & Ours   & 0.50M  & \textbf{1.00T} & \textbf{93.7} & \textbf{5.32E6 (-9.0\%)}  \\ \bottomrule
\end{tabular}
\end{adjustbox}
  \scriptsize
\label{table:p6000}
\vspace{-3mm}
\end{table}

\begin{table}[!tb]
\centering
\renewcommand\arraystretch{1.2}
\caption{Generalizability Evaluation on the Jetson Nano GPU.}
\vspace{1mm}
\begin{adjustbox}{width=0.48\textwidth}
\begin{tabular}{cccccc}
\toprule
{[}JetsonNano{]}                                                              & Method & Params & FLOP/s         & Acc.\%        & Time (ms)               \\ \midrule
\multirow{2}{*}{\begin{tabular}[c]{@{}c@{}}VGG16\\ (CIFAR10)\end{tabular}}    & HRank  & 1.90M  & 35.5G          & \textbf{93.1} & 60.5                    \\ \cline{2-6} 
                                                                              & Ours   & 2.04M  & \textbf{39.4G} & 92.9          & \textbf{48.1 (-20.5\%)} \\ \midrule
\multirow{4}{*}{\begin{tabular}[c]{@{}c@{}}ResNet56\\ (CIFAR10)\end{tabular}} & HRank-1  & 0.48M  & 25.6G          & 93.6          & 82.1                    \\ \cline{2-6} 
                                                                              & Ours   & 0.50M  & \textbf{28.8G} & \textbf{93.7} & \textbf{68.0 (-17.1\%)} \\ \cline{2-6} 
                                                                              & HRank-2  & 0.24M  & 16.7G          & 92.3          & 67.2                    \\ \cline{2-6} 
                                                                              & Ours   & 0.39M  & \textbf{24.3G} & \textbf{92.5} & \textbf{58.1 (-13.3\%)} \\ \bottomrule
\end{tabular}
\end{adjustbox}
\label{table:jetson}
\vspace{-3mm}
\end{table}


\vspace{-2mm}
\section{{Discussion and Conclusion}}
\vspace{-1mm}
Here we discuss the uniqueness and significance of our work and then draw the conclusion.

\textit{This work extends the concept of structured DNN optimization and corresponding granularity.
Rather than taking structured filter configuration as the optimization units, the basic GPU runtime granularity, e.g., waves, also needs to be considered.}


\vspace{-1mm}
\textit{Although this work mainly focuses on GPUs, similar tailing effects also potentially exist in other platforms with massive parallelism.
Therefore, the proposed optimization methodology in this work also has significant potential to be applied for more other computing platforms.}



In summary, our current proposed DNN optimization framework can effectively eliminate the GPU tail effect for DNN execution.
With effectiveness, feasibility, and generalizability well proved, our work shows outstanding DNN latency accuracy trade-offs within various computing scenarios.


\bibliography{example_paper}

\begin{thebibliography}{24}
\providecommand{\natexlab}[1]{#1}
\providecommand{\url}[1]{\texttt{#1}}
\expandafter\ifx\csname urlstyle\endcsname\relax
  \providecommand{\doi}[1]{doi: #1}\else
  \providecommand{\doi}{doi: \begingroup \urlstyle{rm}\Url}\fi

\bibitem[AMD(2020)]{polaris}
AMD.
\newblock \emph{Polaris Architecture | New Graphics Architecture | AMD}, 2020.
\newblock URL \url{https://www.amd.com/technologies/polaris}.

\bibitem[Cai et~al.(2018)Cai, Zhu, and Han]{proxylessnas}
Cai, H., Zhu, L., and Han, S.
\newblock Proxylessnas: Direct neural architecture search on target task and
  hardware.
\newblock \emph{arXiv preprint arXiv:1812.00332}, 2018.

\bibitem[Deng et~al.(2009)Deng, Dong, Socher, Li, Li, and Fei-Fei]{imgnet}
Deng, J., Dong, W., Socher, R., Li, L.-J., Li, K., and Fei-Fei, L.
\newblock Imagenet: A large-scale hierarchical image database.
\newblock In \emph{2009 IEEE conference on computer vision and pattern
  recognition}, pp.\  248--255. Ieee, 2009.

\bibitem[Graves et~al.(2013)Graves, Mohamed, and Hinton]{speech}
Graves, A., Mohamed, A.-r., and Hinton, G.
\newblock Speech recognition with deep recurrent neural networks.
\newblock In \emph{2013 IEEE international conference on acoustics, speech and
  signal processing}, pp.\  6645--6649. IEEE, 2013.

\bibitem[Han et~al.(2015)Han, Mao, and Dally]{hansong}
Han, S., Mao, H., and Dally, W.~J.
\newblock Deep compression: Compressing deep neural networks with pruning,
  trained quantization and huffman coding.
\newblock \emph{arXiv preprint arXiv:1510.00149}, 2015.

\bibitem[He et~al.(2017)He, Zhang, and Sun]{prune1}
He, Y., Zhang, X., and Sun, J.
\newblock Channel pruning for accelerating very deep neural networks.
\newblock In \emph{Proceedings of the IEEE International Conference on Computer
  Vision}, pp.\  1389--1397, 2017.

\bibitem[He et~al.(2018)He, Kang, Dong, Fu, and Yang]{soft}
He, Y., Kang, G., Dong, X., Fu, Y., and Yang, Y.
\newblock Soft filter pruning for accelerating deep convolutional neural
  networks.
\newblock In \emph{Proceedings of the 27th International Joint Conference on
  Artificial Intelligence}, pp.\  2234--2240, 2018.

\bibitem[He et~al.(2019)He, Liu, Wang, Hu, and Yang]{prune2}
He, Y., Liu, P., Wang, Z., Hu, Z., and Yang, Y.
\newblock Filter pruning via geometric median for deep convolutional neural
  networks acceleration.
\newblock In \emph{Proceedings of the IEEE Conference on Computer Vision and
  Pattern Recognition}, pp.\  4340--4349, 2019.

\bibitem[Li et~al.(2016)Li, Kadav, Durdanovic, Samet, and Graf]{lihao}
Li, H., Kadav, A., Durdanovic, I., Samet, H., and Graf, H.~P.
\newblock Pruning filters for efficient convnets.
\newblock \emph{arXiv preprint arXiv:1608.08710}, 2016.

\bibitem[Li et~al.(2019)Li, Zhou, Pan, and Feng]{partial}
Li, X., Zhou, Y., Pan, Z., and Feng, J.
\newblock Partial order pruning: for best speed/accuracy trade-off in neural
  architecture search.
\newblock In \emph{Proceedings of the IEEE Conference on computer vision and
  pattern recognition}, pp.\  9145--9153, 2019.

\bibitem[Lin et~al.(2020)Lin, Ji, Wang, Zhang, Zhang, Tian, and Shao]{hrank}
Lin, M., Ji, R., Wang, Y., Zhang, Y., Zhang, B., Tian, Y., and Shao, L.
\newblock Hrank: Filter pruning using high-rank feature map.
\newblock In \emph{Proceedings of the IEEE/CVF Conference on Computer Vision
  and Pattern Recognition}, pp.\  1529--1538, 2020.

\bibitem[Lin et~al.(2014)Lin, Maire, Belongie, Hays, Perona, Ramanan,
  Doll{\'a}r, and Zitnick]{coco}
Lin, T.-Y., Maire, M., Belongie, S., Hays, J., Perona, P., Ramanan, D.,
  Doll{\'a}r, P., and Zitnick, C.~L.
\newblock Microsoft coco: Common objects in context.
\newblock In \emph{European conference on computer vision}, pp.\  740--755.
  Springer, 2014.

\bibitem[NVIDIA(2020{\natexlab{a}})]{cuda}
NVIDIA.
\newblock \emph{CUDA Programming Guide}, 2020{\natexlab{a}}.
\newblock URL
  \url{https://docs.nvidia.com/cuda/cuda-c-programming-guide/index.html}.

\bibitem[NVIDIA(2020{\natexlab{b}})]{titanv}
NVIDIA.
\newblock \emph{NVIDIA TITAN V | Volta Architecture}, 2020{\natexlab{b}}.
\newblock URL \url{https://www.nvidia.com/en-us/titan/titan-v/}.

\bibitem[Sandler et~al.(2018)Sandler, Howard, Zhu, Zhmoginov, and Chen]{mobnet}
Sandler, M., Howard, A., Zhu, M., Zhmoginov, A., and Chen, L.-C.
\newblock Mobilenetv2: Inverted residuals and linear bottlenecks.
\newblock In \emph{Proceedings of the IEEE conference on computer vision and
  pattern recognition}, pp.\  4510--4520, 2018.

\bibitem[Stamoulis et~al.(2019)Stamoulis, Ding, Wang, Lymberopoulos, Priyantha,
  Liu, and Marculescu]{single_path}
Stamoulis, D., Ding, R., Wang, D., Lymberopoulos, D., Priyantha, B., Liu, J.,
  and Marculescu, D.
\newblock Single-path nas: Designing hardware-efficient convnets in less than 4
  hours.
\newblock In \emph{Joint European Conference on Machine Learning and Knowledge
  Discovery in Databases}, pp.\  481--497. Springer, 2019.

\bibitem[Sze et~al.(2020)Sze, Chen, Yang, and Emer]{vivie}
Sze, V., Chen, Y.-H., Yang, T.-J., and Emer, J.~S.
\newblock Efficient processing of deep neural networks.
\newblock \emph{Synthesis Lectures on Computer Architecture}, 15\penalty0
  (2):\penalty0 1--341, 2020.

\bibitem[Tan \& Le(2019)Tan and Le]{effnet}
Tan, M. and Le, Q.~V.
\newblock Efficientnet: Rethinking model scaling for convolutional neural
  networks.
\newblock \emph{arXiv preprint arXiv:1905.11946}, 2019.

\bibitem[Tan et~al.(2019)Tan, Chen, Pang, Vasudevan, Sandler, Howard, and
  Le]{mnasnet}
Tan, M., Chen, B., Pang, R., Vasudevan, V., Sandler, M., Howard, A., and Le,
  Q.~V.
\newblock Mnasnet: Platform-aware neural architecture search for mobile.
\newblock In \emph{Proceedings of the IEEE Conference on Computer Vision and
  Pattern Recognition}, pp.\  2820--2828, 2019.

\bibitem[Wang et~al.(2019)Wang, Ye, He, Ma, Zhang, Lin, Yuan, Tan, Li, Fan,
  et~al.]{harmful}
Wang, Y., Ye, S., He, Z., Ma, X., Zhang, L., Lin, S., Yuan, G., Tan, S.~H., Li,
  Z., Fan, D., et~al.
\newblock Non-structured dnn weight pruning considered harmful.
\newblock \emph{arXiv preprint arXiv:1907.02124}, 2019.

\bibitem[Wittenbrink et~al.(2011)Wittenbrink, Kilgariff, and Prabhu]{fermi}
Wittenbrink, C.~M., Kilgariff, E., and Prabhu, A.
\newblock Fermi gf100 gpu architecture.
\newblock \emph{IEEE Micro}, 31\penalty0 (2):\penalty0 50--59, 2011.

\bibitem[Wu et~al.(2019)Wu, Dai, Zhang, Wang, Sun, Wu, Tian, Vajda, Jia, and
  Keutzer]{fbnet}
Wu, B., Dai, X., Zhang, P., Wang, Y., Sun, F., Wu, Y., Tian, Y., Vajda, P.,
  Jia, Y., and Keutzer, K.
\newblock Fbnet: Hardware-aware efficient convnet design via differentiable
  neural architecture search.
\newblock In \emph{Proceedings of the IEEE Conference on Computer Vision and
  Pattern Recognition}, pp.\  10734--10742, 2019.

\bibitem[Yang et~al.(2018)Yang, Howard, Chen, Zhang, Go, Sandler, Sze, and
  Adam]{net_adapt}
Yang, T.-J., Howard, A., Chen, B., Zhang, X., Go, A., Sandler, M., Sze, V., and
  Adam, H.
\newblock Netadapt: Platform-aware neural network adaptation for mobile
  applications.
\newblock In \emph{Proceedings of the European Conference on Computer Vision
  (ECCV)}, pp.\  285--300, 2018.

\bibitem[Zhang et~al.(2015)Zhang, Li, Sun, Guan, Xiao, and
  Cong]{time_from_conv1}
Zhang, C., Li, P., Sun, G., Guan, Y., Xiao, B., and Cong, J.
\newblock Optimizing fpga-based accelerator design for deep convolutional
  neural networks.
\newblock In \emph{Proceedings of the 2015 ACM/SIGDA international symposium on
  field-programmable gate arrays}, pp.\  161--170, 2015.

\end{thebibliography}
\bibliographystyle{mlsys2020}

\end{document}